\documentclass{llncs}
\usepackage{graphicx}
\usepackage{amsmath}
\usepackage{amssymb}
\usepackage[ruled, lined, algo2e, linesnumbered]{algorithm2e}
\usepackage[T1]{fontenc}
\usepackage{enumerate}
\usepackage[font=small, format=hang, labelfont=bf]{caption}
\usepackage[]{caption}
\usepackage{subfig}
\usepackage{color}

\begin{document}

%%% Title of contribution
\title{A Fraud Detection Visualization System Utilizing Radial
Drawings and Heat-maps}

%%% Title for running head
\titlerunning{A Fraud Detection Visualization System Utilizing Radial
Drawings and Heat-maps}

%%%% Title for the TOC
\toctitle{A Fraud Detection Visualization System Utilizing Radial
Drawings and Heat-maps}

%%% Author list of contribution

\author{Evmorfia N. Argyriou \inst{1}, A. Symvonis \inst{1} and Vassilis Vassiliou \inst{2}}

%%%% Author list for running head
\authorrunning{E.N. Argyriou, A. Symvonis, V. Vassiliou}

%%%%% List of authors for the TOC
\tocauthor{Evmorfia N. Argyriou, Antonios Symvonis, Vassilis
Vassiliou }

%%%% Abbreviated author list for running head in case of space limitations
\authorrunning{E.N. Argyriou, A. Symvonis, V. Vassiliou}

\institute{%
    School of Applied Mathematical \& Physical Sciences,\\
    National Technical University of Athens, Greece.\\
    \email{$\{$fargyriou,symvonis$\}$@math.ntua.gr}
    \and
    Vodafone Greece.\\
    \email{Vassilis.Vassiliou@vodafone.com}
}

\maketitle

\abstract{We present a prototype system developed in cooperation
with a business organization that combines information visualization
and pattern-matching techniques to detect fraudulent activity by
employees. The system is built upon common fraud patterns searched
while trying to detect occupational fraud suggested by internal
auditors of a business company. The main visualization of the system
consists of a multi-layer radial drawing that represents the
activity of the employees and clients. Each layer represents a
different examined pattern whereas heat-maps indicating suspicious
activity are incorporated in the visualization. The data are first
preprocessed based on a decision tree generated by the examined
patterns and each employee is assigned a value indicating whether or
not there exist indications of fraud. The visualization is presented
as an animation and the employees are visualized one by one
according to their severity values together with their related
clients.}

\section{Introduction}
\label{sec:introduction}

Internal fraud detection in business organizations gains more and
more attention as fraudulent activity appears in ascendant trend
during the last years. Fraud is defined as \emph{the intentional
misuse or abuse of the assets of a company} and may be committed by
employees, clients or other entities~\cite{ACFE12}. Studies on
business fraud schemes show that most of the reported fraud cases
have been committed by trusted associates and this is referred to as
``occupational or employee fraud''. Occupational fraud can be
classified into three main categories: (i)~False or fraudulent
financial statements, (ii)~assets misappropriation and,
(iii)~corruption \cite{ACFE12}. The falsification of financial
statements, produces great loss to a company and is mostly committed
by employees in senior management or executives. Assets
misappropriation is committed by lower-level personnel and due to
the fact that produces insignificant losses at an individual level
it cannot be easily traced by the auditors. Such schemes may
continue for years until fraud is confirmed, producing a huge cost
both to the global economy and to the company. As a result of fraud,
business reputation, company value, and public and client trust are
negatively affected.

\begin{figure*}[h!tb]
  \centering
  \includegraphics[width=.99\textwidth]{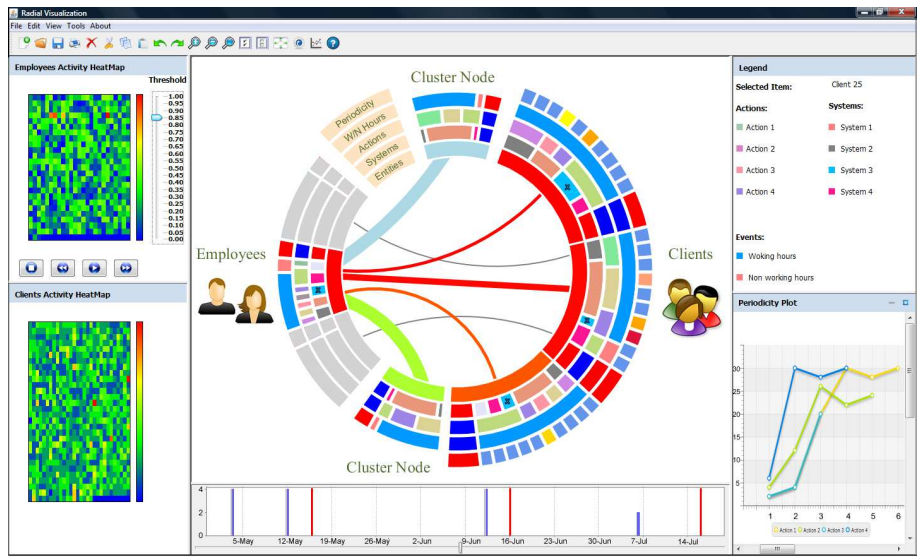}
  \caption{A snapshot of the interface of the system.}
  \label{fig:snapshot}
\end{figure*}

Even though advanced information technology has been incorporated
into organizations to reduce the risk of internal fraud, monitoring
diverse systems that produce textual logs in non-uniform formats is
a time-consuming task. Information visualization can be promising
since it facilitates the quick identification of fraudulent
activity. In this paper, we present a system developed in
cooperation with a business organization that exploits the
advantages of information visualization and pattern recognition to
detect suspicious patterns concerning fraudulent financial
statements in systems in which a pair of entities (employee and
client) are involved. Towards this direction, the system produces a
multi-layer radial drawing (see Figure~\ref{fig:snapshot})
representing the activity of employees and clients along with other
significant information that enable the identification of possible
fraud patterns.

\begin{figure*}[h!tb]
  \centering
  \includegraphics[width=.99\textwidth]{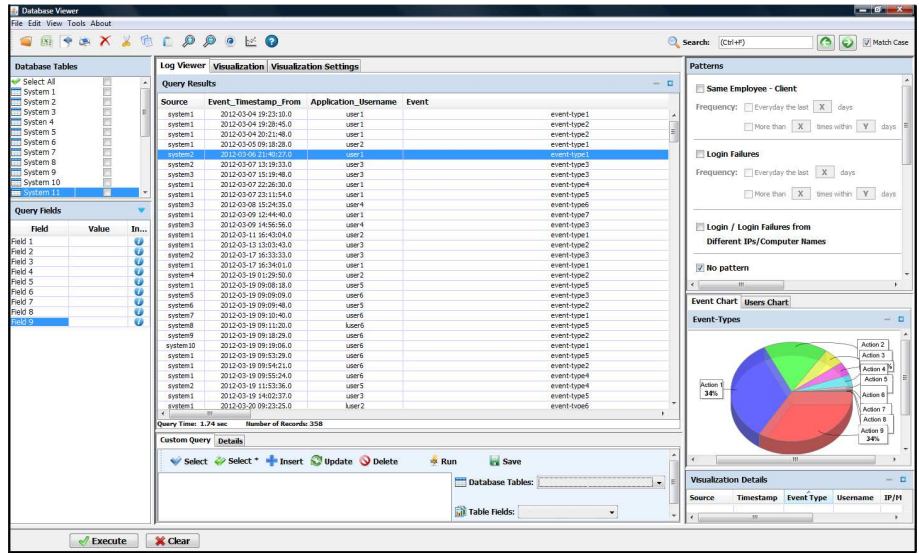}
  \caption{The startup screen of the system. Custom queries can be performed and results are presented in the log viewer.
  The data have been anonymized for security and data privacy reasons.}
  \label{fig:heat-startup-screen}
\end{figure*}

Since occupational fraud schemes are well-hidden in the huge amount
of data, we were seeking for an approach that would present to the
auditor all the recorded events according to their severity. On the
other hand, visualizing large data-sets simultaneously is confusing
and inefficient. For this reason, the system measures the similarity
of the activity of the employees based on fraud detection patterns
(suggested by auditors based on their experience and the framework
of the company on internal fraud risk reduction) and appropriate
heat-maps are generated and incorporated in the system. The produced
visualization is presented as an animation. The system supports
supplementary functionalities such as a database log viewer, export
log mechanisms, storing and post-processing of data, plots and
charts (see Figure~\ref{fig:heat-startup-screen}).

\section{Detection Procedure}
\label{sec:overview}

Fraud detection has been studied enough in the literature. To the
best of our knowledge, there exist only few works oriented
exclusively on occupational fraud detection. Luell~\cite{Lue10}
utilizes data-mining and visualization techniques to detect client
advisor fraud in a financial institution. Eberle and
Holder~\cite{EH09} detect structural anomalies in transactions and
processes propagated by employees using a graph representation.
SynerScope~\cite{Sy11} is an industrial visualization tool for
analyzing ``Big Data'' capable of detecting financial fraud using a
visualization scheme similar to ours representing the billing links
and relations between the company and other entities. The main
difference is that our system is oriented exclusively on
occupational fraud detection based on patterns suggested by auditors
and, thus, it is equipped with a detection mechanism that
preprocesses the data and the visualization conforms to these
patterns. We also utilize animations for the detection of fraud
schemes to avoid cluttering the visualization. A visualization
system based on concentric circles was presented in \cite{AS12}
aiming at identifying periodic events using an algorithm for
periodicity detection. Our system extends the one presented in
\cite{ASS13} that detects periodic patterns that may conceal
occupational fraud in several ways: (i)~The visualization of our
system provides a complete view of all the examined patterns and the
results of the examination on each pattern (in \cite{ASS13}, the
detection procedure was a ``black-box'' determining the order of the
presentation of the clients in a video representing their activity
in which suspicious clients appeared first; partial results of the
detection procedure were illustrated in additional plots and charts,
which hindered the investigation), (ii)~the detection mechanism is
based on a decision tree even though we have incorporated most of
the patterns presented in \cite{ASS13}, (iii)~for periodicity
detection, we apply a variation of the Longest Common Subsequence
algorithm \cite{VHGK03} that tackles noisy data, (iv)~a parallel
coordinates plot has been added to detect unusual employee behavior
(unauthorized access to computers, business systems, etc), (v)~the
system provides a database-viewer to facilitate the investigation
procedure.

Many of the existing publications that deal with fraud detection in
general, use of data-mining techniques \cite{BH02}, \cite{KLSH04},
\cite{PKSG10}. The Financial Crimes Enforcement Network AI
System~\cite{GS95}, \cite{SGW*95} is a system that correlates and
evaluates all reported transactions for indications of money
laundering using pattern-matching techniques. The NASD Regulation
Detection System~\cite{KSH*S98}, \cite{SGS*02} identifies patterns
of violative activity in the Nasdaq stock market by combining
pattern matching and data mining techniques and provides
visualizations of the results. The Link Analysis
Workbench~\cite{WBH*03} searches for criminal activity and terrorism
in noisy and incomplete data also utilizing pattern matching
techniques. Visualization techniques have been also proposed for
financial fraud detection. $3D$-treemaps have been used to monitor
the real-time stock market performance and to identify a stock that
may represent an unusual trading pattern \cite{HLN09}.
``WireVis''~\cite{CLG*08} is a system that provides interactive
visualizations of financial wire transactions and aims to detect
financial fraud. A system that correlates data and discovers complex
networks of potentially illegal financial activities based on
visualization techniques was presented in~\cite{GDLP10}.
``VISFAN''~\cite{DLMP11} has been developed for the visual analysis
of financial activity networks that tries to discover financial
crimes, like money laundering and frauds. ``VIS4AUI''~\cite{DLM12}
is a system that tries to detect money laundering and financial
crimes by collecting financial information related to ongoing bank
relationships and high value transactions. Radial drawings are
widely used for the visualization of large data-sets, especially in
bioinformatics~\cite{CBB*07}, \cite{DM03} and social
networks~\cite{BKD03}, where a large portion of information has to
be visualized simultaneously. This paper is structured as follows:
In Section~\ref{sec:heat-overview}, we describe the detection
procedure. In Section~\ref{sec:heat-description}, we present the
features of the visualization of the
system.Section~\ref{sec:case-study} presents a case-study based on
real data. We conclude in Section~\ref{sec:conclusions} with open
problems and future work.

% ============================================================================
\section{Overview of the Detection Procedure}
\label{sec:heat-overview}
% ============================================================================

As input, the system takes log files from diverse control systems
that are appropriately parsed and stored in a database using a
uniform format. Records generated by systems involving an employee
and a client consist (among other secondary fields) of a time-stamp,
an employee ID, a client ID and an action. An event, say $e$, is
defined as a $4$-tuple $(t,u,c,a)$, where: (i)~$t$ corresponds to
the time-stamp of the occurrence of $e$, (ii)~$u$ corresponds to an
employee, (iii)~$c$ represents a client, and (iv)~$a$ is the action
taken by the employee. For an event $e=(t,u,c,a)$, we say that
client $c$ \emph{is related to} $e$ and is also \emph{related to}
employee $u$. For a pair of employee-client $(u,c)$, an
\emph{event-series} $T_{u,c} = \{e_{(u,c)}^1, e_{(u,c)}^2, \ldots
\}$ is a sequence of events $e_{(u,c)}^i=(t_i,u,c,a_i)$ related to
client $c$ and employee $u$.

\begin{figure}[h!tb]
  \centering
  \includegraphics[width=\textwidth]{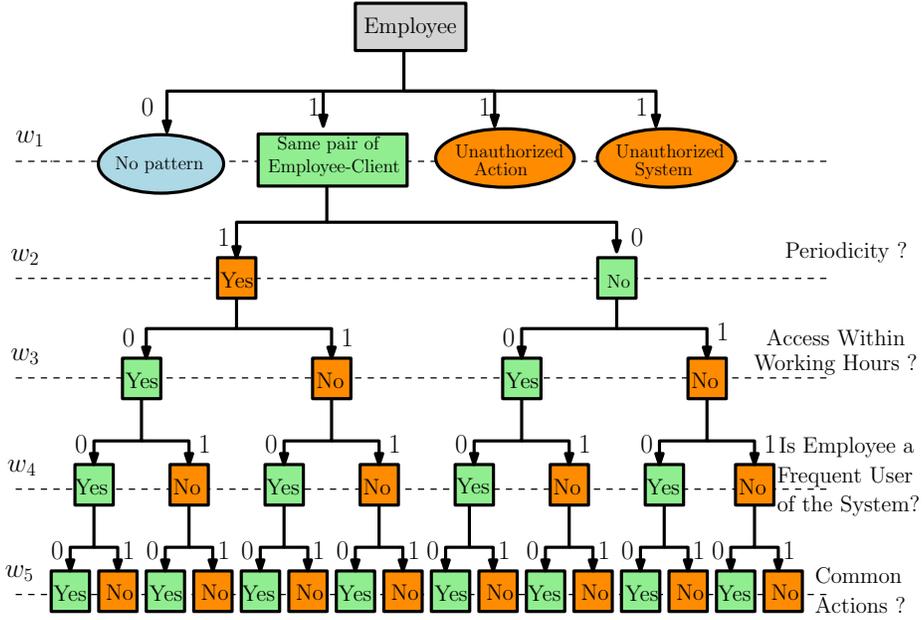}
  \caption{The decision tree based on which the event-series of the employees are assigned a severity value.}
  \label{fig:heat-decision-tree}
\end{figure}

The visualization may be generated either based on the whole
data-set of the database or on queries performed by the auditor in
the startup screen of the system (see
Figure~\ref{fig:heat-startup-screen}). As mentioned above, data have
to be preprocessed before producing the visualization such that
employees with strong indication of fraud are distinguished. For
this reason, for a given employee $u$, the event-series with each
client related to $u$ will be evaluated based on possible fraud
patterns and a value indicating the severity of the related events
(within range $[0,1]$) will be assigned first to the event-series
and then, to the employee. If several fraud patterns are identified,
employee $u$ is assigned the maximum severity value of the already
calculated event-series related to $u$. The evaluation is performed
based on a decision tree generated by the following patterns
suggested by the auditors (see Figure~\ref{fig:heat-decision-tree}):
(i)~There exist more than $X$ events related to employee $u$ and
client $c$ within a time interval of $Y$ days/months where $X,Y$ are
configurable by the auditor (refer to the green rectangle node of
the first layer below the root of the decision tree in
Figure~\ref{fig:heat-decision-tree}), (ii)~the employee has
performed unauthorized actions (based on a list of actions provided
by the auditors) and, (iii)~the employee operated in systems that
she is not authorized to use. In the case where patterns (ii) or
(iii) occur, the event-series of the employee is assigned the
maximum value (i.e., value $1$) such that the employee is definitely
distinguished in the visualization. However, in the case where
pattern (i) occurs, the investigation has to proceed further. The
patterns that are taken into consideration in this case include the
following:
\begin{itemize}
  \item \textbf{Event-series periodicity:} A common pattern while examining
      such fraud schemes is the occurrence of the events in regular time
      basis. For instance, an employee modifies intentionally the account
      of a client every month within the billing cycle of the account and
      more precisely, before its billing date. Assuming that event-series
      $T_{u,c}$ related to employee $u$ and client $c$ is ordered according
      to the time-stamps of the events, the system aims to detect
      similarities between pattern time-series based on a variation of
      Longest Common Subsequence (LCSS) algorithm for time-series \cite{VHGK03} which is
      robust under noisy conditions. The pattern time-series include the
      ideal time-series if the events between the entities appear in time
      intervals that equal exactly to $1,7,15,30$ days and other time-series
      identified in the past as fraud patterns. In the case where
      similarity with any of the above time-series is detected, we consider
      the event-series of the employee to be periodical.
  \item \textbf{Events occurring outside working hours:} Fraudulent
      activities usually occur outside working hours, on weekends, on holidays or at the
      end of the employee's shift. For this reason, if such events occur
      they have to be taken under consideration.
  \item \textbf{Employee frequency in recorded systems:} Each employee
      according to her responsibilities operates in specific business
      systems. If this is not the case, then the employee has to justify
      the recorded event. Also, in several systems such as fraud management
      systems (FMS), it is expected that an employee monitors the activity
      of a suspicious client. Hence, events stemming from these systems
      have to be given smaller weight.
  \item \textbf{Actions taken by the employee:} Similarly to the previous
      case, there exist some actions that an employee is unlikely, but not
      unauthorized to perform since they do not conform to her
      responsibilities.
\end{itemize}

The idea behind the decision tree was to correspond to each layer
one fraud pattern and create a path according to the result of the
examination on each pattern. We consider the importance of patterns
based on their corresponding layer in the decision tree, such that
the higher ones (closer to the root) are more important. Let
$x=[x_1,x_2,x_3,x_4,x_5]$ be the pattern vector examined (e.g.,
$x=[0,0,0,0,0]$ corresponds to non-fraudulent activity) and let
$y=[y_1, \ldots,y_5]$ be the vector resulting from the traversal of
the decision tree from its root to its leaves according to the
evaluation of the events of pair $(u,c)$ on each factor. If the
examination of the events leads to ``Unauthorized action'' or
``Unauthorized system'' the event-series is directly assigned value
$1$. Else, each tree layer $i$ is assigned a weight, say $w_i, i =
1,\ldots,5$, based on the formula presented in \cite{SSE81} such
that dissimilarities between the two vectors that occur at higher
levels of the tree will be more important. For this reason, the
distance between vectors $x,y$, say $d(x,y)$, representing the
dissimilarity by the pattern vector, is calculated by applying the
normalized Weighted Euclidean Distance metric formula
$d(x,y)=\sqrt{\sum_{i=1}^5(x_i-y_i)^2*w_i}/\sqrt{\sum_{i=1} ^5w_i}$.
This value (or the maximum of the already calculated values, if more
than one fraud patterns exist) corresponds to the severity value of
the event-series of employee $u$, which is the value finally
assigned to the employee. Based on these values the system generates
a heat-map representing all employees by rectangular nodes and
gradient colors from blue to red (refer to the upper-left heat-map
of Figure~\ref{fig:snapshot}), such that nodes with color close to
color-red represent employees with strong indications of fraud,
whereas blue colored nodes employees with no suspicion of fraud.
Similarly to the severity calculation of the event-series of
employees, the system assigns also a severity value to clients based
on the above patterns. The only difference is that for a given
client $c$, the severity value is calculated on all the events
related to $c$ (not only the ones that concern a specific employee).
In this manner, a client involved in suspicious activity with two or
more employees will be distinguished.

% ============================================================================
\section{Description of the System}
\label{sec:heat-description}
% ============================================================================

The visualization window\footnote{The reader is suggested to print
the paper in color.} consists of two heat-maps representing the
severity of the activity of employees and clients (refer to the
upper-left and the bottom-left panels of Figure~\ref{fig:snapshot}).
Although a great deal of research has focused on the deficiencies of
rainbow color maps \cite{BT07}, they are still widely used in the
visualizations, since their effectiveness depends on the nature of
the data. In our data-sets, it was necessary to have three colors
(red, green, blue) representing clearly the severity value of the
examined entities (high, medium, low, resp), and thus, we have
chosen this type of color map. We have tried different types of
color maps, but the result was more misleading during the
investigation. The particular selection of colors was not confusing
to the auditors since fraud cases appear rarely in a company which
implies that few red-colored rectangles appear eventually in the
heat-maps.

At startup only the heat-map representing the activity of the
employees is generated. The auditor selects a threshold value that
determines the employees that will be presented based on their
severity values. The visualization is animated and each time an
employee together with her related clients is illustrated. Before an
employee is presented, the heat-map that corresponds to the activity
of related clients is generated.

\begin{figure}[h!tb]
  \centering
  \includegraphics[width=\textwidth]{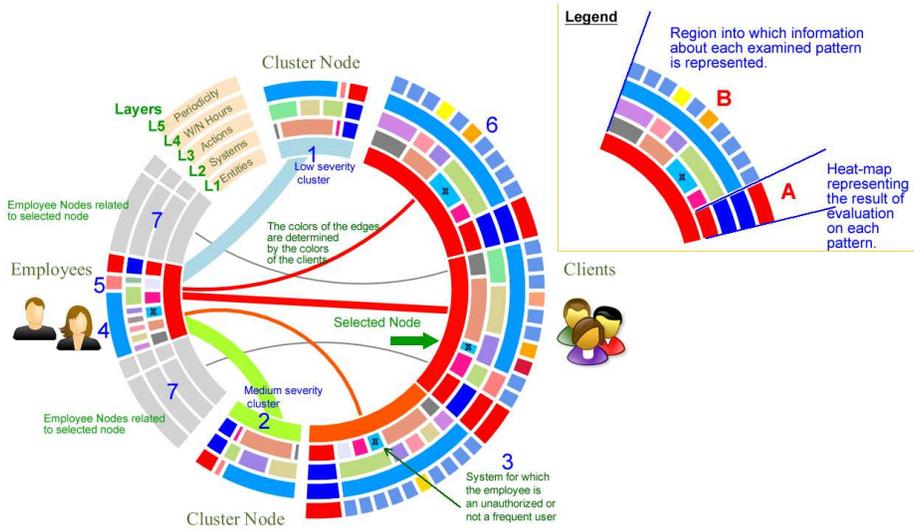}
  \caption{Description of the main-visualization of the system. ``Fat'' edges (unless referring to cluster nodes) and in particular, the red-colored ones may be indications
    of fraud that have to be further examined.}
  \label{fig:heat-description}
\end{figure}

The main visualization of the system is a multi-layer radial drawing
where each layer $L_1,\ldots,L_5$ (see
Figure~\ref{fig:heat-description}) represents a different aspect of
the audit data (systems, actions, within working hours or not,
periodicity) and each circular sector corresponds to an entity
(employee, client or cluster). Then, a graph is generated; its nodes
correspond to each of the above entities, whereas its edges
correspond to the connections among them. The innermost layer of the
visualization (layer $L_1$ of Figure~\ref{fig:heat-description})
accommodates the nodes of the graph (drawn as portions of a ring).
Nodes representing employees are drawn to the left-part of the
visualization where the ones representing their related clients to
the right part. To avoid cluttering the visualization with nodes
representing clients with no indication of fraud, the auditor can
specify thresholds that split clients in one or two clusters
(low-severity cluster and/or medium-severity cluster) according to
their severity values. These nodes are accommodated on top and to
the bottom part of the visualization. The color of the nodes (apart
from the ones representing clusters and the gray-colored ones that
will be explained later) follows the color of the corresponding
entities in the heat-maps. The light-blue (green, resp) colored
cluster-node corresponds to the low (medium, resp) severity cluster
(refer to reference points $1$ and $2$ of
Figure~\ref{fig:heat-description}, resp.). Regarding the edges of
the drawing, the system supports either circular arc edges or
straight-line edges. The thickness of an edge is proportional to the
number of connections between the employee and the client while its
color is determined by the color of the client. In a fraud context,
``fat'' edges (unless referring to cluster nodes) and in particular,
the red-colored ones will be indications of fraud that have to be
further examined.

Subsequent layers (i.e., layers $L_2-L_5$) represent the patterns
described in Section~\ref{sec:heat-overview} and are split into two
regions $A$ and $B$ (refer to Figure~\ref{fig:heat-description}).
Region $A$ represents a heat-map indicating the result of the
examination of the entity in the specific pattern. Red color
indicates identification of suspicious pattern. Region $B$
illustrates information about each corresponding examined pattern.
In layer $L_2$, the different business systems related to the each
of the entities are represented. Each such system is characterized
by a specific color and occupies space proportional to the
corresponding aggregate percentage of use by the entity. For each
employee, this percentage is calculated based on the aggregated
percentage of use on all clients that are currently drawn in the
visualization, whereas for each client based on the percentage of
use by the employee currently visualized (unless more than one
employees related to a client are drawn simultaneously in the
visualization). Similarly, for cluster nodes the aggregated
percentage of use for all clients that belong to the cluster is
calculated. Systems for which the employee is an unauthorized or not
a frequent user are marked by an $X$ (refer to reference point $3$
of Figure~\ref{fig:heat-description}).

Layer $L_3$ corresponds to the actions reported for each entity, and
are drawn in a similar manner as the ones in layer $L_2$. Again
unauthorized or suspicious actions are marked with an $X$. Layer
$L_4$ represents the percentage of events that occur within or
outside working hours. The light-blue colored parts represent events
occurring within working hours, whereas the light-red colored parts
indicate the existence of events occurring outside working hours
(e.g., see reference points $4$ and $5$ of
Figure~\ref{fig:heat-description}, resp). For each client node there
exists an additional layer (refer to $L_5$ of
Figure~\ref{fig:heat-description}, e.g., see reference point $6$)
that indicates whether or not the event-series of the client is
periodical. The event-series is compared with the pattern
time-series currently stored in the system and a heat-map is
generated indicating the degree of similarity with each pattern.
Again, light-red colors indicate suspicious cases.

Regarding the investigation procedure, as already mentioned, the
auditor specifies a threshold and the employees with assigned
severity value above the threshold are presented in the
visualization one by one, together with their related clients. The
auditor is able to start, pause or stop the video and process the
visualization. In the case where a client node is selected,
additional employees related to the client can be added to the
visualization which facilitates the possible identification of two
or more employees that may cooperate in committing fraud (the case
where an employee node is selected is treated similarly). In
Figure~\ref{fig:heat-description}, gray colored nodes (see reference
point $7$) represent nodes added during post-processing when a
client node is selected (refer to the node pointed by the green
arrow of Figure~\ref{fig:heat-description}). In the case where more
than one node representing employees exist simultaneously in the
visualization, the auditor is able to select one of them and add the
related clients to the visualization. Gray-color is utilized for
non-selected employee nodes together with their edges and related
clients (if they are not related also to the selected employee) to
avoid distracting the auditor.

Since it is possible to switch between employees during the
investigation, if the post-processing of a case is completed, the
animation resumes from the last visualized employee before pausing
the animation. In the case where a cluster node is selected, the
corresponding rectangles in the heat-map representing clients are
marked with an $X$ such that they can be added (if desired) to the
visualization. However, the system permits a specific number of
additions of employee or client nodes in order to avoid cluttering
the visualization area. If this number is exceeded, the system
optionally is able to produce a visualization where only the
inner-most layer of the radial drawing (i.e., the one corresponding
to the clients and employees) is drawn along with the relations
between them. Again, the width of the edges is proportional to the
number of events that relate two entities. In the case where further
investigation is needed the auditor selects the desired node (which
becomes larger) and the other layers of the radial drawing (i.e.,
$L_2-L_5$) that correspond to the particular node appear in the
visualization. In this manner, the system is able to visualize
simultaneously a larger set of entities and reveal the relations
between them. However, this may slow down the investigation process
and for this reason, we adopted both the animation approach and the
simultaneous visualization of all layers for each node.

The ordering of the nodes representing the employees is performed
according to their severity value (the more suspicious nodes are
presented first in the animation). The clients appear in arbitrary
order since no crossings between edges connecting a particular
employee with her related clients can exist. The only crossings that
may occur are caused by gray-colored employees related to clients
already visualized and since these edges are also gray-colored, they
do not confuse the auditor (see Figure~\ref{fig:heat-description}).
The gray-colored employees that are added in the visualization are
placed either on the top or to the bottom of the already placed
employee-nodes to retain the relative positions of the already
placed nodes. We also have chosen not to apply crossing minimization
heuristics since, the addition of new nodes to the visualization may
imply a rearrangement of the positions of the already placed nodes,
disturbing the underlying mental map.

\begin{figure}[h!tb]
  \centering
  \includegraphics[width=\textwidth]{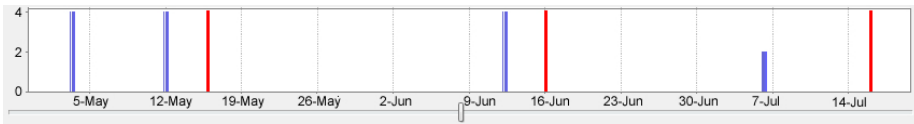}
  \caption{A time-line plot representing the event-series for the selected pair of entities.}
  \label{fig:timeline-plot}
\end{figure}

\begin{figure}[h!tb]
  \centering
  \includegraphics[width=.4\textwidth]{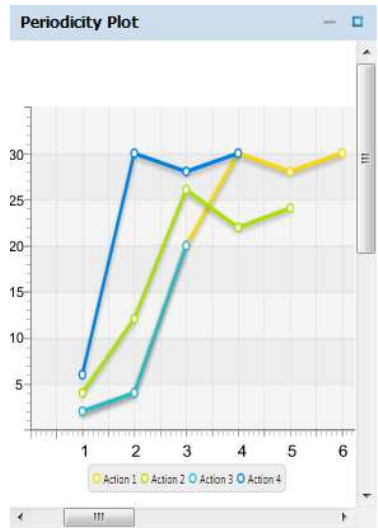}
  \caption{A periodicity-plot that presents all reported actions by distinct series for the selected pair of entities.}
  \label{fig:periodicity-plot}
\end{figure}

Figure~\ref{fig:timeline-plot} accommodates a time-line plot
representing the event-series for the selected pair of entities
(refer to the blue series) where the $x$-axis corresponds to the
date of the occurrence of each event and the $y$-axis to the number
of events occurred during the specific date. The red drawn columns
represent the billing dates of the account of the specific client.
This plot facilitates the identification of possible periodic
activity especially close to the billing date of the account of the
client. In employee fraud schemes, it is also possible that the
event-series related to a pair of entities is periodical only based
on a specific action. For this reason, we have incorporated in the
system a second plot (refer to Panel $\#5$ of
Figure~\ref{fig:periodicity-plot}) that presents all reported
actions by distinct series. In this plot, the $y-$axis corresponds
to the day of the month that event $x$ related to a specific action
occurred (e.g., the first event related to an action that occurred
on the $15$th day of a month will be drawn on point $(1,15)$. For
this case, we ignore multiple occurrences of events related to the
same action occurred the same day of the month. In the case where
periodicity occurs for a specific action, the corresponding series
will have part (or the whole series) almost parallel to $x-$axis.

\begin{figure}[t]
  \centering
  \includegraphics[width=\textwidth]{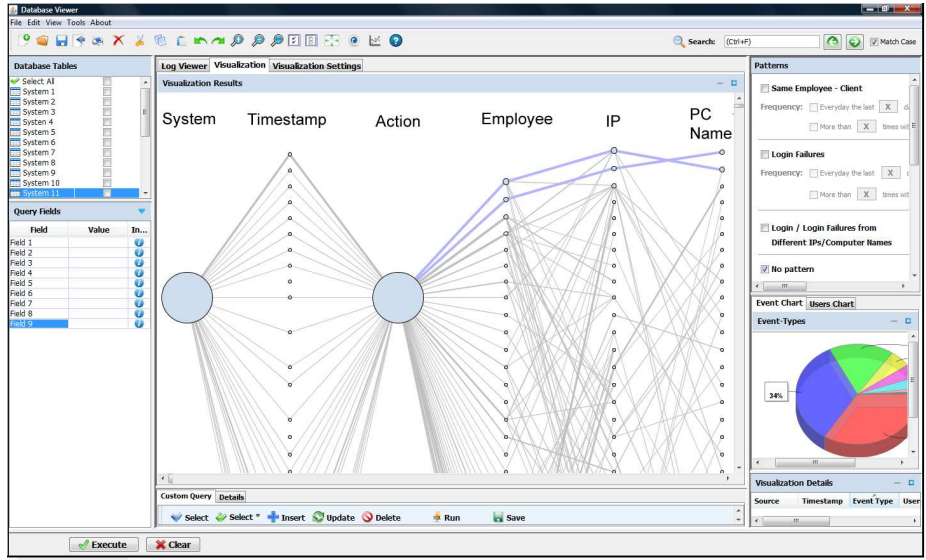}
  \caption{A parallel coordinates plot to monitor failed login attempts in a specific system.}
  \label{fig:heat-parallel-plot}
\end{figure}

The system supports mechanisms to detect patterns of unusual
employee behavior such as unauthorized access to computers, business
systems and accounts of employees or clients by producing a parallel
coordinates plot (see Figure~\ref{fig:heat-parallel-plot}). Each
record consists of an employee, a time-stamp, an IP indicating the
address of the employee's computer, a computer name and an action
(i.e., login, login failure, etc). The size of the nodes and the
edges is proportional to the number of their occurrences in the
database. The nodes on each layer are ordered by their number of
occurrences in the data-set. The patterns used in this scenario
include the following: (i)~More than $X$ failed login attempts
within a time interval of $Y$ days/months, where $X,Y$ are
configurable by the auditor, and (ii)~Login attempts or failed login
attempts occurring from different IPs and/or computer names. The
visualization of Figure~\ref{fig:heat-description} can also be
adapted to this scenario -either with or without the preprocessing
step (since the number of employees and actions is manageable in a
radial drawing)- by substituting the nodes representing the clients
by nodes that represent actions.

% ============================================================================
\section{Case Study}
\label{sec:case-study}
% ============================================================================

In this section, we present the results of the evaluation of the
system on real data-sets stemming from two control systems of a
telecommunication company. All data provided to us were anonymous
for security and data privacy reasons. The data-set consists of
approximately $180.637$ entries lying within a time interval of six
months. The data-set consists of $710$ distinct employees and
$83.030$ distinct clients. In the data-set, $66.2\%$ of clients had
only one occurrence, $31.6\%$ between $2$ and $5$, $1.4\%$ between
$5$ and $10$ while the remaining ones (i.e., $0.8\%$) had more than
$10$ occurrences. The auditors have also included a set of entries
corresponding to a ``fictional'' fraud case scenario where an
employee modifies the account of a client. However, we were not
communicated any information regarding the billing date of the
accounts of the clients.

\begin{figure}[h!t]
  \centering
  \includegraphics[width=.4\textwidth]{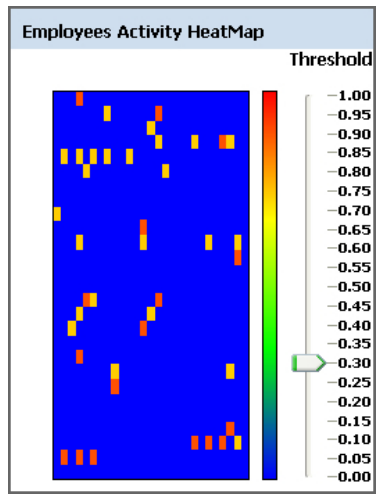}
  \caption{A heat-map indicating the severity values for the employees participated in the case study.}
  \label{fig:heat-map}
\end{figure}

Since one of the data-sets stems from a fraud management system, it
is expected that reoccurring activity between the same pair of
employee and client will occur (a ``suspicious'' client reported by
a fraud management system is expected to be supervised by an
employee). For this reason, we concentrated our study in identifying
pairs of employees and clients that appear to have more that $10$
related events. The system identified $41$ employees ($5.8\%$ of the
total number of employees) that were related with the same client
with more than $10$ events (refer to the orange and red colored
rectangles of the heat-map of Figure~\ref{fig:heat-map}).

For each of the above employees we had to calculate the similarity
of the event-series with pattern time-series (see
Section~\ref{sec:heat-overview}) in order to detect periodicity. In
particular, the auditors were interested for periodic events that
occur monthly (i.e., periodicity of $28$ up to $30$ days). Thus, we
could consider only the pattern time-series (refer to
Section~\ref{sec:heat-overview}) that corresponds to monthly
activity. However, we decided to calculate the severity values based
on all pattern time-series in order to distinguish any reoccurring
activity and decide afterwards through the visualization if
suspicious activity really exists. Even though, this approach
creates more false positives that have to be investigated, it
ensures that other possible periodic events will not omitted. Among
these $41$ employees that were related with the same client with
more than $10$ events, $17$ of them ($2.4\%$ of the total number of
employees) appear to have periodic activity and events occurring
outside working hours (refer to the red-colored rectangles of
Figure~\ref{fig:heat-map}), while the remaining ones ($24$
employees; 3.4\% of the total number of employees) had events
occurring outside working hours (refer to the orange-colored
rectangles of Figure~\ref{fig:heat-map}). Also, no employee had
performed unauthorized access to systems or had used unauthorized
actions, whereas only one employee had used non common actions. The
results were communicated to the auditors who investigated whether
there exist real indications of fraud.

\begin{figure}[h!t]
  \centering
  \includegraphics[width=\textwidth]{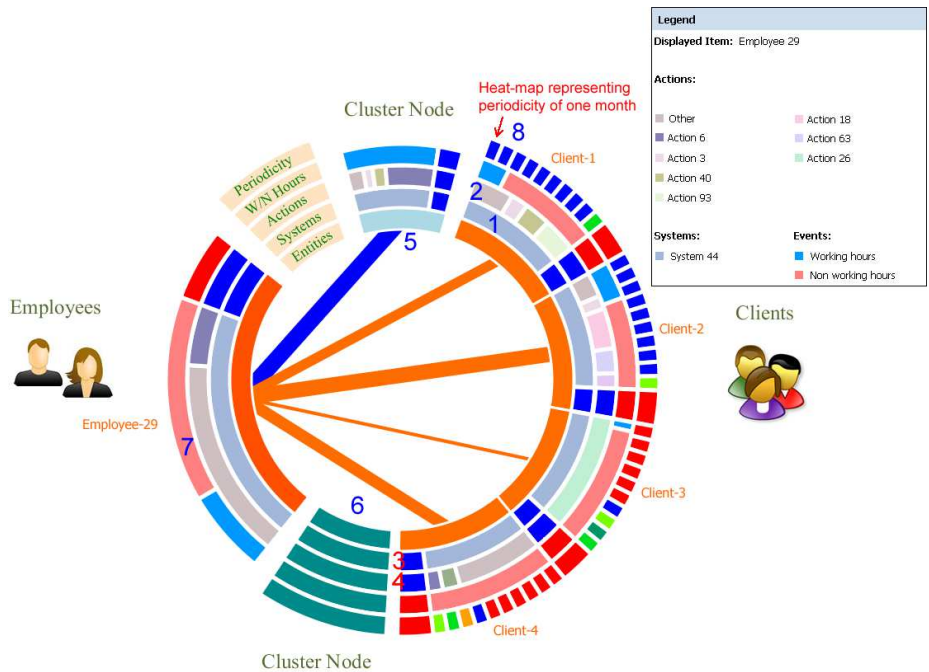}
  \caption{A frame of the animation indicating the activity of highly-ranked Employee-$26$.}
  \label{fig:userY}
\end{figure}

In the following, we present some of the first frames of the
animation in order to describe the investigation procedure.
Figure~\ref{fig:userY} indicates a highly-ranked employee (i.e.,
Employee-$29$). Employee-$29$ was related to four clients, i.e.,
Client-$1$, Client-$2$, Client-$3$ and Client-$4$, more than $10$
times using one system for which the employee was a frequent user
and common actions (see reference points $1$ and $2$ of
Figure~\ref{fig:userY}; the parts representing the systems and the
actions are not marked by an X, which implies that the user was
frequent for the specific systems or had performed common actions,
and the corresponding heat-maps are blue-colored; see reference
points $3$ and $4$ of Figure~\ref{fig:userY}). All other clients
related to Employee-$29$, where placed in the low-severity cluster,
whereas no client was placed in the medium-severity cluster (see
reference points $5$ and $6$ of Figure~\ref{fig:userY}, resp.). Most
of the events related to Employee-$26$ (see reference point $7$ of
Figure~\ref{fig:userY}). The layer of the visualization which
accommodates the actions that Employee-$26$ has performed, is split
in two parts representing ``Action 6'' and ``Other''. In the
``Other'' action, we have clustered actions whose percentage of use
was too small ($<0.1\%$) to be visualized. The problem was caused by
the large number of possible actions in this particular system which
does not permit the simultaneous visualization of all actions.
However, if one of the clustered actions is not common for a
particular employee, the red-colored heat-map in the corresponding
layer and the $X$ marking in the part representing the ``Other''
action would reveal the problem to the auditor.

We will now proceed to describe the investigation procedure for each
of the four clients. Client-$1$ and Client-$2$ have the majority of
the events that relate them with Employee-$29$ occurring outside
working hours (refer to the corresponding layer of Client-$1$ and
Client-$2$). However, they appear to have a small similarity (less
than $50\%$) with only one fraud pattern time-series. Note that, the
pattern time-series that corresponds to periodicity of one month
corresponds to the first heat-map of the each periodicity layer (see
e.g., reference point $8$ of Figure~\ref{fig:userY}). Since the
auditors were interested in events that occur monthly, these clients
were not considered suspicious. Regarding Client-$3$ and Client-$4$,
they have also the majority of the events that relate them with
Employee-$29$ occurring outside working hours but, they appear to
have strong periodic activity. In particular, there exist six
heat-maps indicating strong similarities with pattern time-series
including the one that represents monthly periodic activity. For
this reason, these clients have to be further investigated.

\begin{figure}[h!t]
  \centering
  \includegraphics[width=\textwidth]{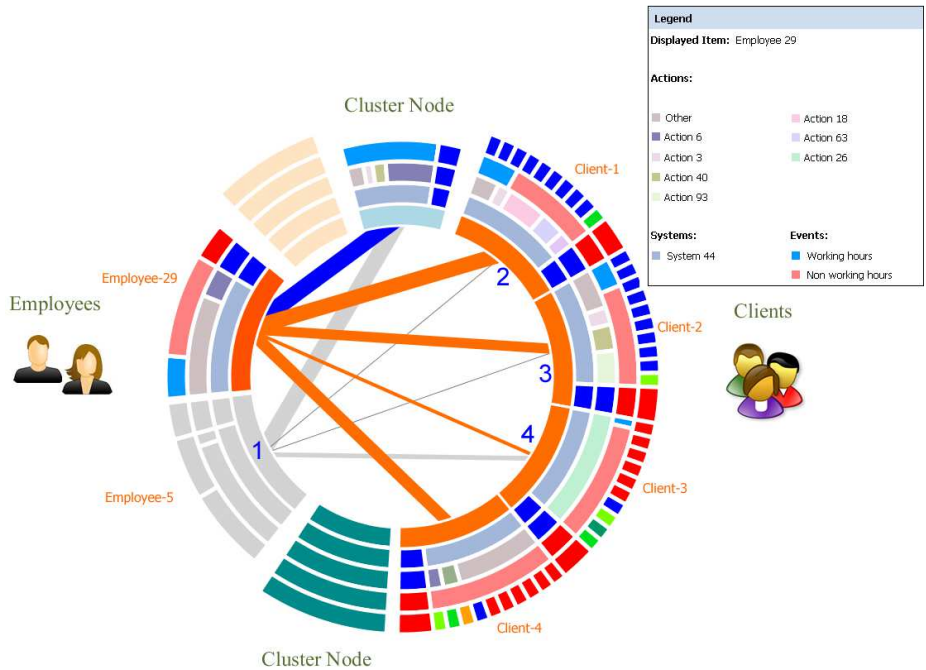}
  \caption{All employees related to Client-$3$ are added in the visualization.
  The gray-colored node corresponds to the new employee added (Employee-$5$) who surprisingly is related to almost all high-ranked clients of Employee-$29$.}
  \label{fig:userY-2}
\end{figure}

\begin{figure}[h!tb]
  \centering
  \hfill
  \begin{minipage}[b]{\textwidth}
     \centering
     \subfloat[\label{fig:userY-timeline}{}]
     {\includegraphics[width=.9\textwidth]{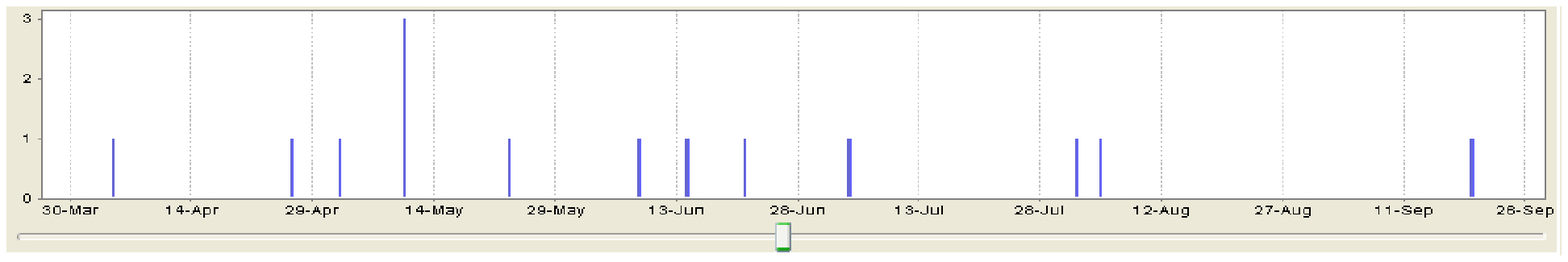}}
  \end{minipage}
  \vfill
  \begin{minipage}[b]{\textwidth}
     \centering
     \subfloat[\label{fig:userY-periodicity}{}]
     {\includegraphics[width=.6\textwidth]{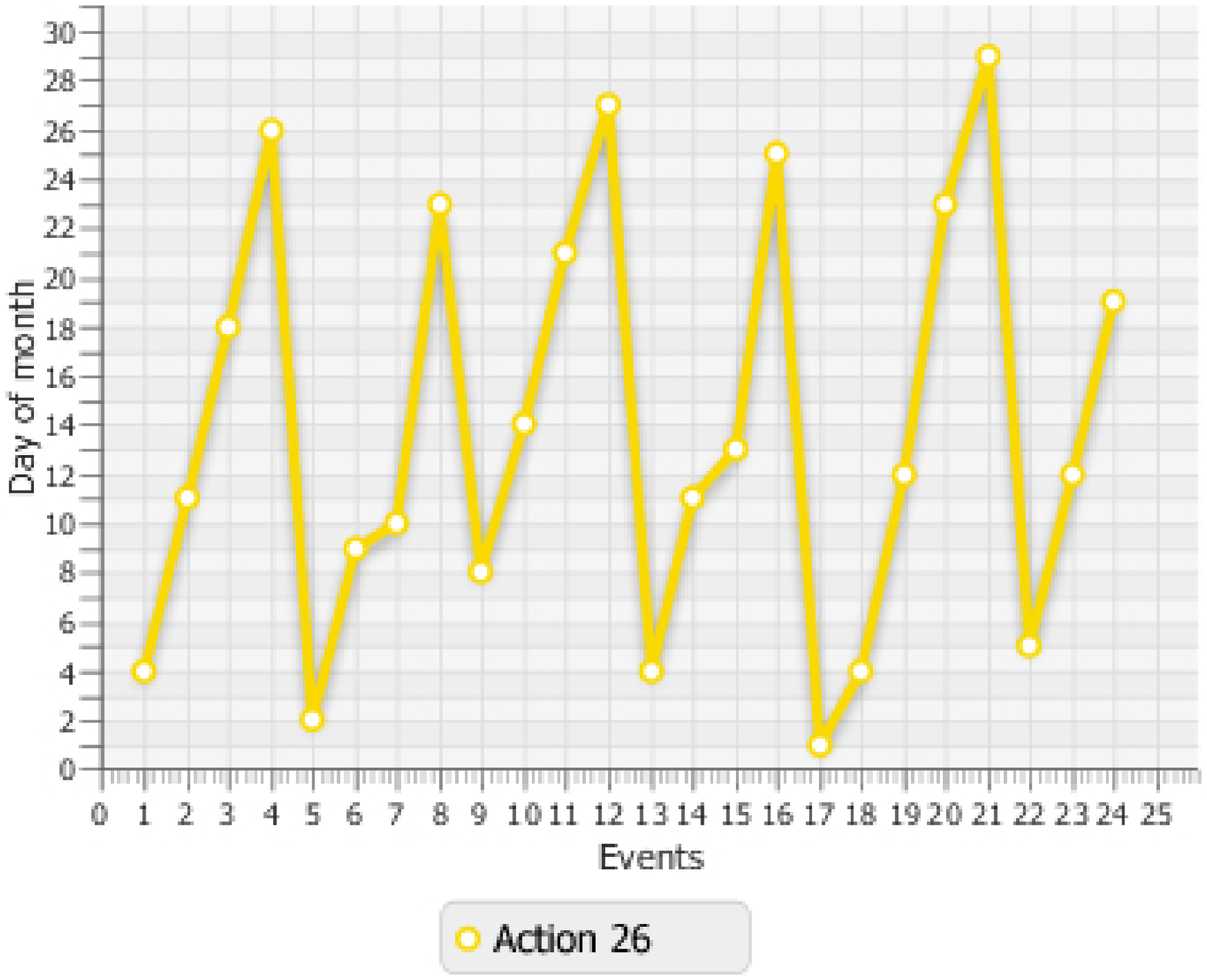}}
  \end{minipage}
  \hfill
  \caption{(i)~The time-line plot for Employee-$40$ and Client-$6$ indicating an obvious monthly activity.
  (ii)~A plot illustrating the periodicity of performed actions for Employee-$40$ and Client-$6$.}
  \label{fig:supplementary-visualizations-for-y}
\end{figure}

Client-$4$ was related to no other employee (when we selected the
node that corresponds to Client-$4$, no other employee was added to
the visualization). However, when we selected Client-$3$, a new
employee, referred to as Employee-$5$ (see the gray-colored node of
the visualization; reference point 1 of Figure~\ref{fig:userY-2}),
was added to the visualization. The interesting thing was that
Employee-$5$ was related with the same high-ranked clients as
Employee-$29$ (except for Client-$4$; see reference points $2,3$ and
$4$ of Figure~\ref{fig:userY-2}). There exist two possible
explanations for this scenario. Either two employees are both
responsible for monitoring the activity of the clients or these
employees are accomplice to fraud. Thus, the investigation has to
proceed further. The time-line plot of
Figure~\ref{fig:userY-timeline} represents the time-stamps of the
events relating Employee-$29$ and Client-$3$. Obviously, there
exists a continuous activity between the two entities. However, this
activity according to the auditors resembles more to monitoring
activity rather than to a fraud pattern. This assumption is also
reinforced by Figure~\ref{fig:userY-periodicity} which represents
all reported actions for the selected pair of entities. In
particular, only one action is performed by Employee-$29$ towards
Client-$3$ and according to the auditors this action is part of a
monitoring procedure. Studying, in a similar manner, the time-line
plot and the periodicity plot for Client-$4$, the auditors claimed
that there existed no indications of fraud (the recorded actions
were also part of a monitoring procedure).

%\begin{figure}[h!t]
%  \centering
%  \includegraphics[width=\textwidth]{images/chapter-9/usery-timeline}
%  \caption{The time-line plot for Employee-$40$ and Client-$6$ indicating an obvious monthly activity.}
%  \label{fig:userY-timeline}
%\end{figure}
%
%\begin{figure}[h!t]
%  \centering
%  \includegraphics[width=.5\textwidth]{images/chapter-9/usery-periodicity}
%  \caption{A plot illustrating the periodicity of performed actions for Employee-$40$ and Client-$6$.}
%  \label{fig:userY-periodicity}
%\end{figure}

\begin{figure}[h!]
  \centering
  \includegraphics[width=\textwidth]{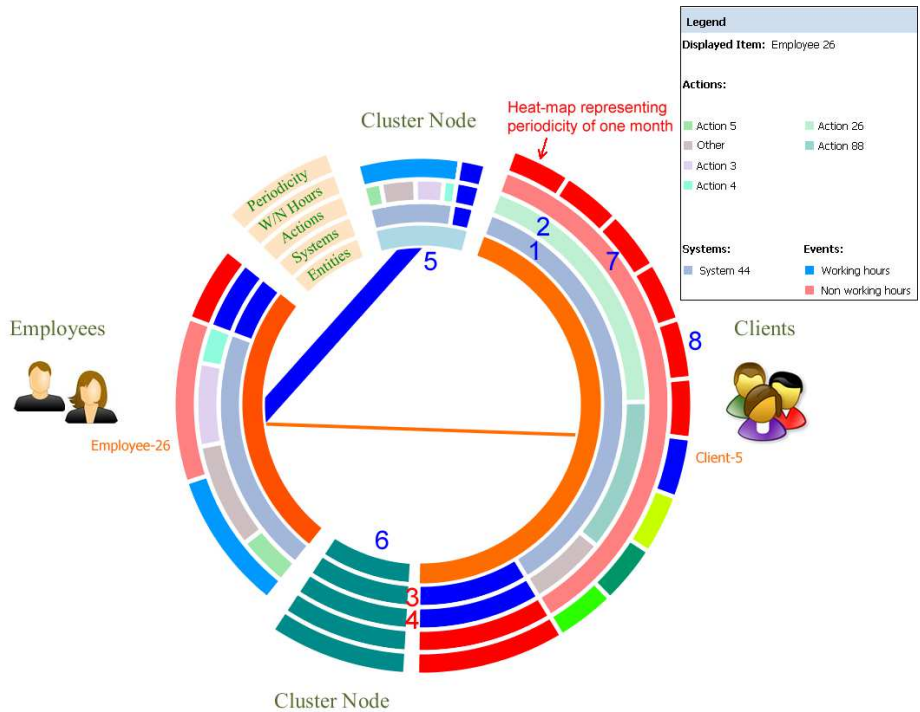}
  \caption{A frame of the animation indicating the activity of a Employee-$26$.}
  \label{fig:userX}
\end{figure}

\begin{figure}[h!]
  \centering
  \includegraphics[width=.9\textwidth]{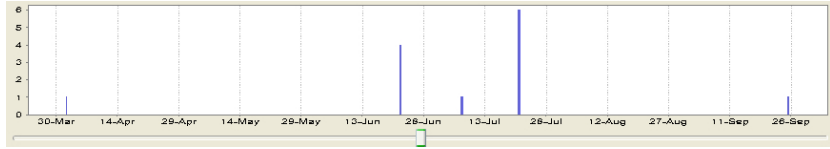}
  \caption{The corresponding time-line plot for Employee-$26$ and Client-$5$.}
  \label{fig:userx-timeline}
\end{figure}

Figure~\ref{fig:userX} depicts another highly-ranked employee (i.e.,
Employee-$26$). Employee-$26$ was involved only with Client-$5$,
more than $10$ times using one system for which the employee was a
frequent user and common actions (see reference points $1$, $2$ $3$
and $4$ of Figure~\ref{fig:userX}). Again, all other clients related
to Employee-$26$, where placed in the low-severity cluster, whereas
no client was placed in the medium-severity cluster (see reference
points $5$ and $6$ of Figure~\ref{fig:userX}, resp.). Also,
Client-$5$ was related only to Employee-$26$ (when we selected the
node that corresponds to Client-$5$, no other employee was added to
the visualization). However, almost all events that relate the two
entities appear outside the working hours (see reference point $7$
of Figure~\ref{fig:userX}). The visualization of
Figure~\ref{fig:userX} also indicates a strong periodic activity
(refer to the heat-maps in the last layer of the part of the
visualization that corresponds to Client-$5$; see reference point
$8$). In the next step, we examined the time-line plot for the two
entities (refer to Figure~\ref{fig:userx-timeline}). One can
distinguish a monthly periodic activity between June and September
(i.e., the actual dates are 22/6, 21/7, 25/9), even with small gaps
(no entries in August) and some noisy data (i.e., 7/7). The auditors
that examined the case determined that there were no indications of
fraud in this particular case mostly because of the actions
performed by the employee which were again common monitoring
actions. Client-$5$ was reported by a fraud management system as a
suspicious client and thus, Employee-$26$ was monitored her in
regular time basis.

\begin{figure}[h!t]
  \centering
  \includegraphics[width=\textwidth]{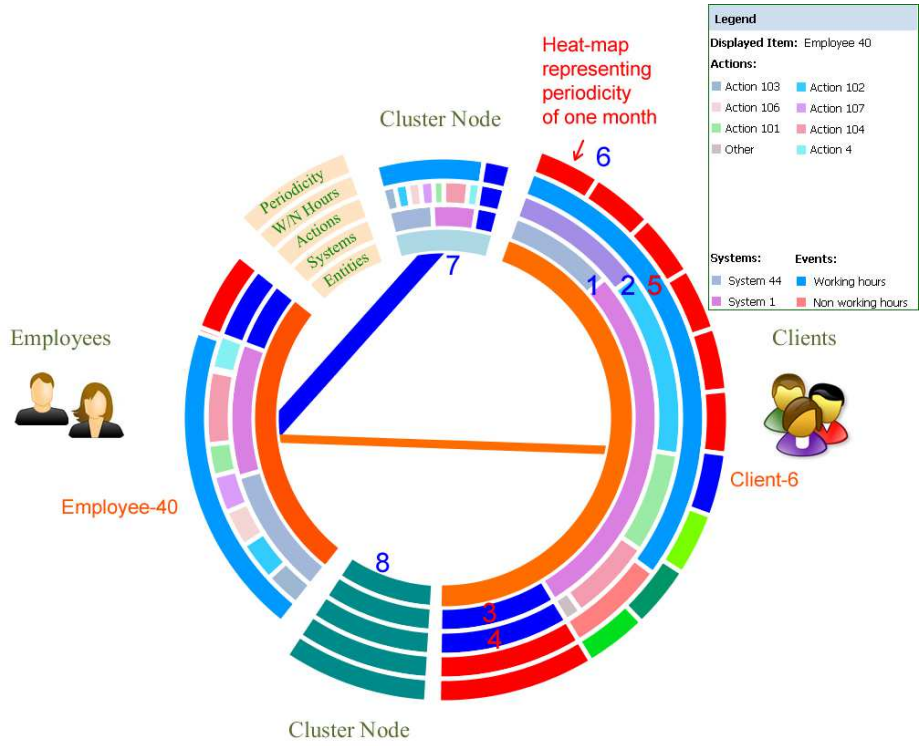}
  \caption{A frame of the animation illustrating the activity of Employee-$40$. This case corresponds to a fictional fraud case scenario.}
  \label{fig:fraud-user}
\end{figure}

\begin{figure}[h!t]
  \centering
  \includegraphics[width=.9\textwidth]{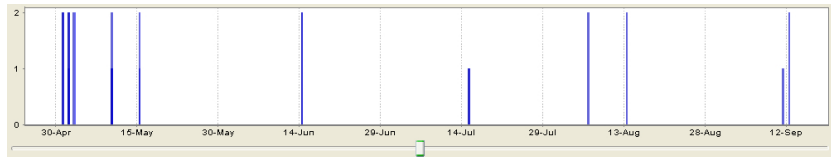}
  \caption{The time-line plot for Employee-$40$ and Client-$6$ indicating an obvious monthly activity.}
  \label{fig:fraud-user-timeline}
\end{figure}

\begin{figure}[h!t]
  \centering
  \includegraphics[width=.5\textwidth]{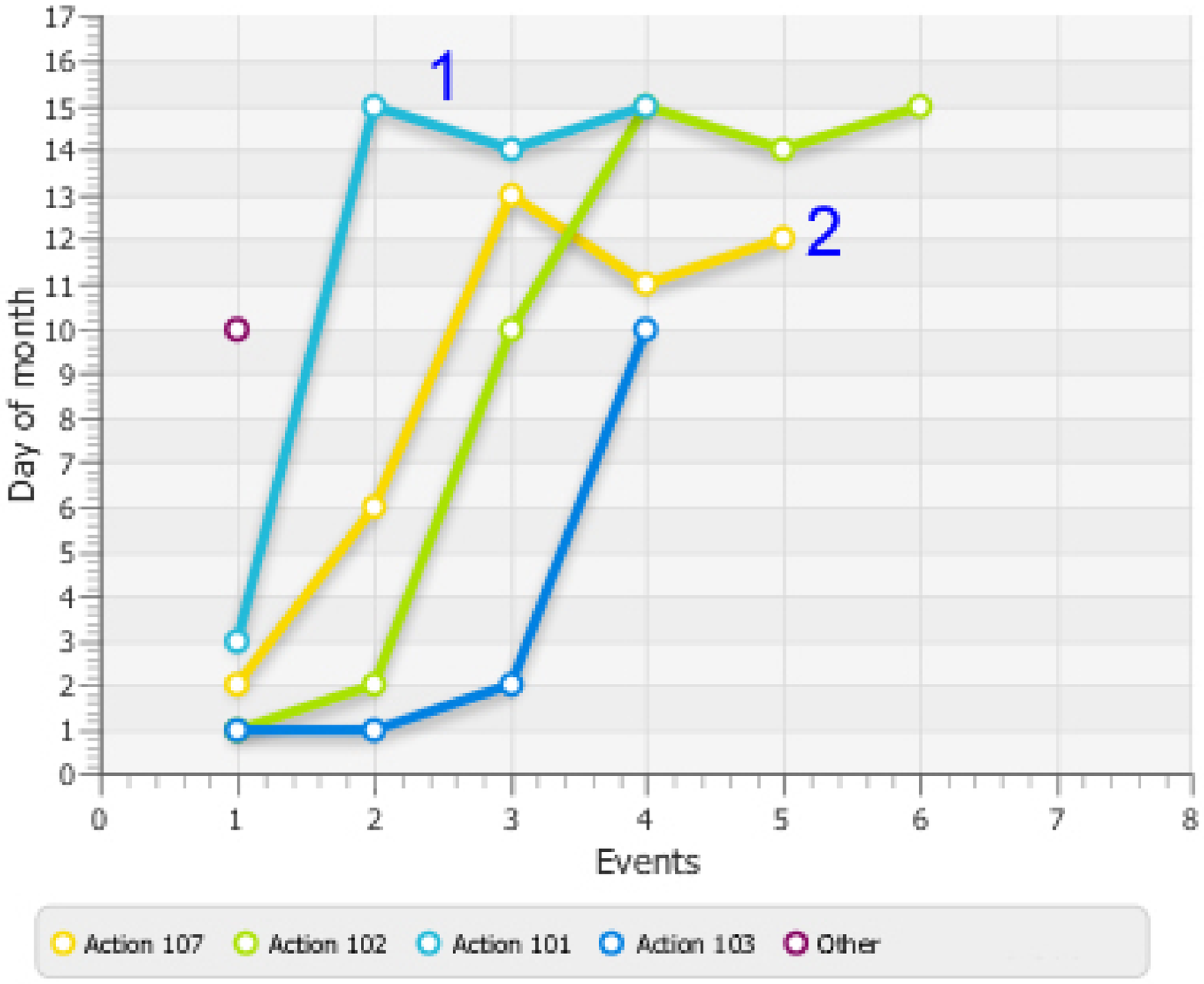}
  \caption{A plot indicating the periodic pattern for the performed actions of Employee-$40$ and Client-$6$.}
  \label{fig:fraud-user-periodicity}
\end{figure}

Figure~\ref{fig:fraud-user} illustrates the fictional fraud case
which was added by the auditors. In this scenario, Employee-$40$ is
related to Client-$6$ using two business systems for which she is a
frequent user and common actions (see reference points $1,2,3$ and
$4$ of Figure~\ref{fig:fraud-user}). All the events occur within
working hours (see reference point $5$ of
Figure~\ref{fig:fraud-user}). However, the event-series of
Employee-$40$ appears to have strong similarity with six fraud
pattern time-series, including the one that corresponds to
periodicity of one month, which is represented by the first heat-map
of the each periodicity layer (see reference point $6$ of
Figure~\ref{fig:fraud-user}).Also, Client-$6$ was not related to any
other employee (when selected, no other employee was added to the
visualization). All other clients related to Employee-$40$, where
placed in the low-severity cluster, whereas no client was placed in
the medium-severity cluster (see reference points $7$ and $8$ of
Figure~\ref{fig:fraud-user}, resp.). Even though, this visualization
resembles a lot to the one of Employee-$26$ (see
Figure~\ref{fig:userX}), the time-line plot (see
Figure~\ref{fig:fraud-user-timeline}) and the periodicity plot (see
Figure~\ref{fig:fraud-user-periodicity}) explain why this case is
considered as fraud. The first suspicions according to the auditors,
are raised by the fact that there exists activity between the two
entities stemming from two business systems (they take also into
consideration the type of the systems; an information that was not
communicated to us in full detail).

The auditors explained to us that the time-line plot (refer to
Figure~\ref{fig:fraud-user-timeline}) matches to a fraud case
scenario according to which there exists some activity at the
beginning (between April - May) with no specific periodicity and
then, appears periodic activity (from May to September). In the
first time interval, the fraudster is trying to plan and organize
her fraud by performing a number of actions. Once the fraud is
organized, only periodic actions are required. Another suspicious
fact in this case is that the events from May to September occur
close to the same dates of each month (from 10th to 15th). Of
course, there exist some ``noisy'' data that have to be excluded in
order to understand the fraud pattern. These may have been caused
either on purpose to cover up the fraud or were part of the duties
of the employee.

The above assumption is reinforced by the plot of
Figure~\ref{fig:fraud-user-periodicity} which reveals the periodic
occurrence of each performed action. For instance, ``Action 101''
(see reference point $1$) appears to have periodicity around the
15th day of the month from its second occurrence and later. Also,
``Action 107'' (see reference point $2$) appears to have periodicity
around the 11th-13th day of the month from its third occurrence and
later. In particular, the vast majority of events are recorded
between the 10th and the 15th day of the month. We could be more
convinced that this case consists fraud if we knew exactly the
billing cycle of the account of the client.

In a similar manner, the frames of the animation illustrating the
other highly-ranked employees were investigated. In particular, we
were given more attention to the $17$ frames of the animation
containing periodic events. Since the data-sets provided for the
case analysis were sensitive, we were not communicated many details
about the final results of the investigation. Fortunately for the
company, the only real evidence of fraud existed in the fictional
data added by the auditors. However, the auditors had not identified
all these cases while examining the data-sets manually and they had
to make an additional investigation for them.

% ============================================================================
\section{\uppercase{Conclusions and Future Work}}
\label{sec:conclusions}
% ============================================================================

We presented an integrated fraud management visualization system
that aims to identify patterns that may conceal occupational fraud
through a combination of pattern recognition and visualization. Our
work opens several aspects for future work such as incorporation of
more fraud patterns, use of more statistical methods and, extension
of the system in order to identify more complicated fraud schemes
(client fraud, telecommunication fraud, etc.) in a wider variety of
business systems.

\section*{\uppercase{Acknowledgements}}

\noindent The work of Evmorfia N. Argyriou has been co-financed by
the European Union (European Social Fund - ESF) and Greek national
funds through the Operational Program "Education and Lifelong
Learning" of the National Strategic Reference Framework (NSRF) -
Research Funding Program: Heracleitus II. Investing in knowledge
society through the European Social Fund.

\vfill
\bibliographystyle{abbrv}
{\small
%\bibliography{references}

\begin{thebibliography}{10}

\bibitem{ASS13}
E.~N. Argyriou, A.~Sotiraki, and A.~Symvonis.
\newblock Occupational fraud detection through visualization.
\newblock In {\em ISI}, pages 4--7, 2013.

\bibitem{AS12}
E.~N. Argyriou and A.~Symvonis.
\newblock Detecting periodicity in serial data through visualization.
\newblock In {\em ISVC}, volume 7432, pages 295--304, 2012.

\bibitem{ACFE12}
{A}ssociation of {C}ertified~{F}raud {E}xaminers.
\newblock {\em Report to the Nation on Occupational Fraud and Abuse}, 2012.

\bibitem{BH02}
R.~J. Bolton and D.~J. Hand.
\newblock Statistical fraud detection: A review.
\newblock {\em Statistical Science}, 17:2002, 2002.

\bibitem{BT07}
D.~Borland and R.~M. Taylor~II.
\newblock Rainbow color map (still) considered harmful.
\newblock {\em IEEE Comput. Graph. Appl.}, 27(2):14--17, 2007.

\bibitem{BKD03}
U.~Brandes, P.~Kenis, and D.~Wagner.
\newblock Communicating centrality in policy network drawings.
\newblock {\em IEEE Transactions on Visualization and Computer Graphics},
  9(2):241--253, 2003.

\bibitem{CLG*08}
R.~Chang, A.~Lee, M.~Ghoniem, R.~Kosara, and W.~Ribarsky.
\newblock Scalable and interactive visual analysis of financial wire
  transactions for fraud detection.
\newblock {\em Information Visualization}, 7(1):63--76, 2008.

\bibitem{CBB*07}
F.~Chevenet, C.~Brun, A.-L. Banuls, B.~Jacq, and R.~Christen.
\newblock Treedyn: towards dynamic graphics and annotations for analyses of
  trees.
\newblock {\em BMC Bioinformatics}, 7(1):1--9, 2007.

\bibitem{DLM12}
W.~Didimo, G.~Liotta, and F.~Montecchiani.
\newblock Vis4aui: Visual analysis of banking activity networks.
\newblock In {\em GRAPP/IVAPP}, pages 799--802, 2012.

\bibitem{DLMP11}
W.~Didimo, G.~Liotta, F.~Montecchiani, and P.~Palladino.
\newblock An advanced network visualization system for financial crime
  detection.
\newblock In {\em PacificVis}, pages 203--210, 2011.

\bibitem{DM03}
S.~N. Dorogovtsev and J.~F.~F. Mendes.
\newblock {\em Evolution of Networks: From Biological Nets to the Internet and
  WWW (Physics)}.
\newblock Oxford University Press, Inc., New York, NY, USA, 2003.

\bibitem{EH09}
W.~Eberle and L.~B. Holder.
\newblock Mining for insider threats in business transactions and processes.
\newblock In {\em CIDM}, pages 163--170, 2009.

\bibitem{GDLP10}
E.~D. Giacomo, W.~Didimo, G.~Liotta, and P.~Palladino.
\newblock Visual analysis of financial crimes.
\newblock In {\em AVI}, pages 393--394, 2010.

\bibitem{GS95}
H.~G. Goldberg and T.~E. Senator.
\newblock Restructuring databases for knowledge discovery by consolidation and
  link formation.
\newblock In {\em KDD}, pages 136--141, 1995.

\bibitem{HLN09}
M.~L. Huang, J.~Liang, and Q.~V. Nguyen.
\newblock A visualization approach for frauds detection in financial market.
\newblock IV '09, pages 197--202, 2009.

\bibitem{KSH*S98}
J.~D. Kirkland, T.~E. Senator, J.~J. Hayden, T.~Dybala, H.~G.
Goldberg, and
  P.~Shyr.
\newblock The nasd regulation advanced detection system (ads).
\newblock In {\em AAAI '98/IAAI '98}, pages 1055--1062, 1998.

\bibitem{KLSH04}
Y.~Kou, C.-T. Lu, S.~Sirwongwattana, and Y.-P. Huang.
\newblock Survey of fraud detection techniques.
\newblock In {\em Networking, Sensing and Control, 2004 IEEE Int. Conf.},
  volume~2, pages 749--754, 2004.

\bibitem{Lue10}
J.~Luell.
\newblock {\em Employee fraud detection under real world conditions}.
\newblock PhD thesis, 2010.

\bibitem{PKSG10}
C.~Phua, V.~C.~S. Lee, K.~Smith-Miles, and R.~W. Gayler.
\newblock A comprehensive survey of data mining-based fraud detection research.
\newblock {\em CoRR}, abs/1009.6119, 2010.

\bibitem{SGS*02}
T.~E. Senator, H.~G. Goldberg, P.~Shyr, S.~Bennett, S.~Donoho, and
C.~Lovell.
\newblock chapter The NASD regulation advanced detection system: integrating
  data mining and visualization for break detection in the NASDAQ stock market,
  pages 363--371. 2002.

\bibitem{SGW*95}
T.~E. Senator, H.~G. Goldberg, J.~Wooton, M.~A. Cottini, A.~F.~U.
Khan, C.~D.
  Klinger, W.~M. Llamas, M.~P. Marrone, and R.~W.~H. Wong.
\newblock The financial crimes enforcement network ai system (fais) identifying
  potential money laundering from reports of large cash transactions.
\newblock {\em AI Magazine}, 16(4):21--39, 1995.

\bibitem{SSE81}
W.~G. Stillwell, D.~A. Seaver, and W.~Edwards.
\newblock A comparison of weight approximation techniques in multiattribute
  utility decision making.
\newblock {\em Organ. Behavior and Human Performance}, 28(1):62 -- 77, 1981.

\bibitem{Sy11}
SynerScope.
\newblock 2011.
\newblock http://www.synerscope.com/.

\bibitem{VHGK03}
M.~Vlachos, M.~Hadjieleftheriou, D.~Gunopulos, and E.~Keogh.
\newblock Indexing multi-dimensional time-series with support for multiple
  distance measures.
\newblock In {\em ACM SIGKDD int. conf. on Knowledge discovery and data
  mining}, KDD '03, pages 216--225, 2003.

\bibitem{WBH*03}
M.~Wolverton, P.~Berry, I.~Harrison, J.~Lowrance, D.~Morley,
A.~Rodriguez,
  E.~Ruspini, and J.~Thomere.
\newblock Law: A workbench for approximate pattern matching in relational data.
\newblock In {\em IAAI}, pages 143--150, 2003.

\end{thebibliography}

}

\end{document}